\documentclass[twocolumn,aps,superscriptaddress,showpacs]{revtex4}

\usepackage{amssymb}
\usepackage{amsmath}
\usepackage{graphicx}
\usepackage[normalem]{ulem}
\usepackage[dvips]{color}
\usepackage{multirow}

\def\es0{$E_{sym}(\rho_0)$~}

\def\us0{$U_{sym}(\rho_0,k_F)$~}

\def\l0{$L(\rho_0)$~}

\begin{document}

\title{Revisit of the neutron/proton ratio puzzle in intermediate-energy heavy-ion collisions}
\author{Hai-Yun Kong}
\affiliation{Shanghai Institute of Applied Physics, Chinese Academy
of Sciences, Shanghai 201800, China}
\affiliation{University of Chinese Academy of Sciences, Beijing 100049, China}
\author{Yin Xia}
\affiliation{Shanghai Institute of Applied Physics, Chinese Academy
of Sciences, Shanghai 201800, China}
\affiliation{University of Chinese Academy of Sciences, Beijing 100049, China}
\author{Jun Xu\footnote{corresponding author: xujun@sinap.ac.cn}}
\affiliation{Shanghai Institute of Applied Physics, Chinese Academy
of Sciences, Shanghai 201800, China}
\author{Lie-Wen Chen}
\affiliation{Department of Physics and
Astronomy and Shanghai Key Laboratory for Particle Physics and
Cosmology, Shanghai Jiao Tong University, Shanghai 200240, China}
\affiliation{Center of Theoretical Nuclear Physics, National
Laboratory of Heavy Ion Accelerator, Lanzhou 730000, China}
\author{Bao-An Li}
\affiliation{Department of Physics and
Astronomy, Texas A$\&$M University-Commerce, Commerce, TX
75429-3011, USA}
\author{Yu-Gang Ma}
\affiliation{Shanghai Institute of Applied Physics, Chinese Academy
of Sciences, Shanghai 201800, China}
\affiliation{Shanghai Tech University, Shanghai 200031, China}

\date{\today}

\begin{abstract}
Incorporating a newly improved isospin- and momentum-dependent interaction
in the isospin-dependent Boltzmann-Uehling-Uhlenbeck
transport model IBUU11, we have investigated relative effects of the density dependence of nuclear
symmetry energy $E_{sym}(\rho)$ and the neutron-proton effective mass splitting $m^*_n-m^*_p$ on the neutron/proton ratio of free nucleons and those in light clusters.
It is found that the $m^*_n-m^*_p$ has a relatively stronger effect than the $E_{sym}(\rho)$ and the assumption of $m^*_n\leq m^*_p$ leads to a higher neutron/proton ratio.
Moreover, this finding is independent of the in-medium nucleon-nucleon cross sections used.  However, results of our calculations using the
$E_{sym}(\rho)$ and $m^*_n-m^*_p$ both within their current uncertainty ranges are all too low compared to the recent NSCL/MSU double neutron/proton ratio data from central
$^{124}$Sn+$^{124}$Sn and $^{112}$Sn+$^{112}$Sn collisions at 50 and 120 MeV/u, thus calling for new mechanisms to explain the puzzlingly high
neutron/proton ratio observed in the experiments.
\end{abstract}


\maketitle


To pin down the density dependence of nuclear symmetry energy $E_{sym}(\rho)$ has long been a
major challenge for both nuclear physics and astrophysics. While much progress has been made
in the past decade, many interesting issues remain to be resolved~\cite{Bar05,Ste05,Lat07,Li08}.
A larger symmetry energy generally corresponds to a more repulsive (attractive) underlying
symmetry potential $U_{sym}$ for neutrons (protons) in neutron-rich nuclear matter as they are linearly proportional to each other according to the Hugenholtz-Van Hove theorem~\cite{hug} or the Bruckner theory~\cite{bru64,Dab73}, see, e.g., Refs.~\cite{XuC10,Rchen,xuli2} for the explicit relationship between $E_{sym}(\rho)$ and $U_{sym}$.
On the other hand, the in-medium nucleon effective mass describes to the first order effects due to the non-locality of the underlying nuclear interactions and the Pauli exchange effects in
many-fermion systems~\cite{jamo}.
It can be calculated from the momentum dependence of the single-particle potential in non-relativistic models or the Schr\"{o}dinger-equivalent potential in relativistic models.
The nucleon effective mass is related to many interesting problems in both nuclear physics and astrophysics~\cite{Jeu76,Cooperstein85,Bethe90,Farine01}. It has further
been found that the neutrons and protons may have different effective masses in neutron-rich matter due to the momentum dependence of the symmetry (isovector) potential.
However, calculations within different models using various interactions, e.g., the Brueckner-Hartree-Fock approach~\cite{Dalen05}, the relativistic
mean-field model~\cite{Che07}, and the Skyrme-Hartree-Fock calculation~\cite{Sto03,OLi11}, predict rather different values for the neutron-proton effective mass splitting $m^*_{n-p}\equiv m^*_n-m^*_p$.
Thus, currently there is no consensus as to whether the $m^*_{n-p}$ is negative, zero, or positive. However, the value of $m^*_{n-p}$ affects significantly
isospin-sensitive observables in heavy-ion collisions~\cite{LiBA04,Che04,Riz05,Li05,Gio10,Fen11,Zha14,Xie14} as well as thermal and transport properties of neutron-rich matter~\cite{Beh11,ImMDI,Xu15}. It also has important ramifications in astrophysics~\cite{Mei07}. For instance, the equilibrium neutron/proton ratio in the primordial nucleosynthesis is determined by $(n/p)_{eq}=e^{-m^*_{n-p}/T}$ in the early ($\geq$ 1ms) universe when the temperature $T$ was high ($\geq$ 3 MeV)~\cite{Ste06}.

Recently, analyses of the free neutron/proton double ratio from central $^{124}$Sn+$^{124}$Sn and $^{112}$Sn+$^{112}$Sn collisions at 50 and 120 MeV/u at the NSCL/MSU seem to indicate that protons have a slightly larger effective mass than neutrons based on comparisons with calculations within an improved quantum molecular dynamics model using Skyrme interactions~\cite{Cou14}. The preferred Skyrme interaction SLy4~\cite{Cha98} is being widely used in describing the ground state properties and excitations of neutron-rich nuclei. However, the applicability of Skyrme interactions is restricted, in general, by the nuclear structure calculations and the
small amplitude nuclear motions due to the incorrect energy behavior of nucleon-nucleus isovector optical potential in comparison with that extracted from experiments~\cite{Jeu91,Rap79,Pat76,Kon03,XuC10,Li13,XHL13,Li14}. This situation clearly calls for more theoretical studies with different transport models and examinations of various model ingredients. In fact, it was known that the isospin-dependent Boltzmann-Uehling-Uhlenbeck (IBUU) transport model using a momentum-dependent symmetry potential corresponding to $m^*_n>m^*_p$ \cite{Li06} under-predicts the old NSCL double neutron/proton data~\cite{NSCL06} no matter how the density-dependence of symmetry energy and the in-medium nucleon-nucleon cross sections are adjusted. Thus, to understand the puzzlingly high neutron/proton double ratio, the relative effects of the symmetry energy and the neutron-proton effective mass splitting should be studied within the same model.
We have recently incorporated a newly improved isospin- and momentum-dependent interaction (ImMDI)~\cite{ImMDI} based on the Gogny force in the IBUU11 transport model. The original Gogny interaction is known to give an asymptotic value of the isoscalar potential at saturation density less than that extracted from optical model analyses of nucleon-nucleus scattering data. We removed this drawback and introduced a new parameter to adjust easily the values of the $m^*_{n-p}$ besides those varyingly the magnitude and density dependence of the $E_{sym}(\rho)$. In this Brief Report, we revisit the neutron/proton puzzle. We found that indeed the $m^*_n-m^*_p$ has a relatively stronger effect than the $E_{sym}(\rho)$ and the assumption of $m^*_n\leq m^*_p$ leads to a higher neutron/proton ratio. However, the puzzle remains as our calculations using the $E_{sym}(\rho)$ and $m^*_n-m^*_p$ both within their current uncertainty ranges still under-predict significantly the NSCL/MSU data.


The ImMDI interaction is developed from the MDI interaction, which
has a similar functional form as the Gogny effective interaction
while replacing the Gaussian-type finite-range term with a Yukawa
form~\cite{Das03,Xu11}. The potential energy density in asymmetric
nuclear matter from the MDI interaction or the ImMDI interaction is
expressed as~\cite{Das03}
\begin{eqnarray}
V(\rho ,\delta ) &=&\frac{A_{u}\rho _{n}\rho _{p}}{\rho _{0}}+\frac{A_{l}}{%
2\rho _{0}}(\rho _{n}^{2}+\rho _{p}^{2})+\frac{B}{\sigma
+1}\frac{\rho
^{\sigma +1}}{\rho _{0}^{\sigma }}  \notag \\
&\times &(1-x\delta ^{2})+\frac{1}{\rho _{0}}\sum_{\tau ,\tau
^{\prime
}}C_{\tau ,\tau ^{\prime }}  \notag \\
&\times &\int \int d^{3}pd^{3}p^{\prime }\frac{f_{\tau }(\vec{r}, \vec{p}%
)f_{\tau ^{\prime }}(\vec{r}, \vec{p}^{\prime })}{1+(\vec{p}-\vec{p}^{\prime
})^{2}/\Lambda ^{2}}. \label{MDIV}
\end{eqnarray}%
In the above, $\rho _{n}$ and $\rho _{p}$ are respectively the
neutron and proton density, and $\rho =\rho _{n}+\rho _{p}$ is the
total density. $\rho _{0}$ is the saturation density, and $\delta
=(\rho _{n}-\rho _{p})/\rho$ is the isospin asymmetry. $\tau=1(-1)$
denoting neutrons (protons) is the isospin index, and $f_{\tau
}(\vec{r}, \vec{p})$ is the phase-space distribution function. The
single-particle potential from the mean-field approximation depends
on the density $\rho$ and isospin asymmetry $\delta$ of the nuclear
medium as well as the isospin $\tau$ and momentum $\vec{p}$ of the
nucleon~\cite{Das03}
\begin{eqnarray}
U_\tau(\rho ,\delta ,\vec{p}) &=&A_{u}\frac{\rho _{-\tau }}{\rho _{0}}%
+A_{l}\frac{\rho _{\tau }}{\rho _{0}}  \notag \\
&+&B\left(\frac{\rho }{\rho _{0}}\right)^{\sigma }(1-x\delta ^{2})-4\tau x\frac{B}{%
\sigma +1}\frac{\rho ^{\sigma -1}}{\rho _{0}^{\sigma }}\delta \rho
_{-\tau }
\notag \\
&+&\frac{2C_{\tau,\tau}}{\rho _{0}}\int d^{3}p^{\prime }\frac{f_{\tau }(%
\vec{r}, \vec{p}^{\prime })}{1+(\vec{p}-\vec{p}^{\prime })^{2}/\Lambda ^{2}}
\notag \\
&+&\frac{2C_{\tau,-\tau}}{\rho _{0}}\int d^{3}p^{\prime }\frac{f_{-\tau }(%
\vec{r}, \vec{p}^{\prime })}{1+(\vec{p}-\vec{p}^{\prime })^{2}/\Lambda ^{2}}.
\label{MDIU}
\end{eqnarray}%

Comparing with the MDI interaction, the ImMDI interaction has been
improved mainly in two aspects~\cite{ImMDI}. First, the high-momentum
part of the nucleon isoscalar mean-field potential has been refitted
to reproduce the optical potential extracted from the proton-nucleus
scattering experimental data up to the nucleon kinetic energy of
about 1 GeV. Second, besides the $x$ parameter, which was previously
used to mimic the density dependence of the symmetry energy by
adjusting the relative contributions of different spin-isospin channels of
the density-dependent interaction, another two parameters $y$ and
$z$ are introduced to vary respectively the isospin splitting of the
nucleon effective mass and the value of the symmetry energy at
saturation density. The parameters $y$ and $z$ enter the functional
form through
\begin{eqnarray}
A_{l}(x,y)&=&A_{0} + y + x\frac{2B}{\sigma +1},  \label{AlImMDI}\\
A_{u}(x,y)&=&A_{0} - y - x\frac{2B}{\sigma +1}, \label{AuImMDI}\\
C_{\tau,\tau}(y,z)&=&C_{l0} - \frac{2(y-2z)p^2_{f0}}{\Lambda^2\ln [(4 p^2_{f0} + \Lambda^2)/\Lambda^2]}, \label{ClImMDI}\\
C_{\tau,-\tau}(y,z)&=&C_{u0} + \frac{2(y-2z)p^2_{f0}}{\Lambda^2\ln
[(4 p^2_{f0} + \Lambda^2)/\Lambda^2]}, \label{CuImMDI}
\end{eqnarray}
where $p_{f0}=\hbar(3\pi^{2}\rho_0/2)^{1/3}$ is the nucleon Fermi
momentum in symmetric matter at saturation density. The
values of $A_{0}$, $C_{u0}$, $C_{l0}$, $B$, $\sigma$, and $\Lambda$
are fixed by six empirical constraints at $x=0$, $y=0$, and $z=0$,
i.e., the saturation density $\rho_{0}=0.16$ fm$^{-3}$, the binding
energy $E_{0}(\rho_{0})=-16$ MeV, the incompressibility $K_{0}=230$
MeV, the isoscalar effective mass $m^{*}_{s,0}=0.7m$, the symmetry
energy $E_{sym}(\rho_{0})=32.5$ MeV, and the isoscalar potential at
infinitely large momentum $U_{0,\infty}=75$ MeV, and the values of
the corresponding parameters are $A_{0}=-66.963$ MeV,
$C_{u0}=-99.7017$ MeV, $C_{l0}=-60.4860$ MeV, $B=141.963$ MeV,
$\sigma=1.26521$, and $\Lambda=2.42401p_{f0}$. 

\begin{figure}[h]
\centerline{\includegraphics[scale=0.8]{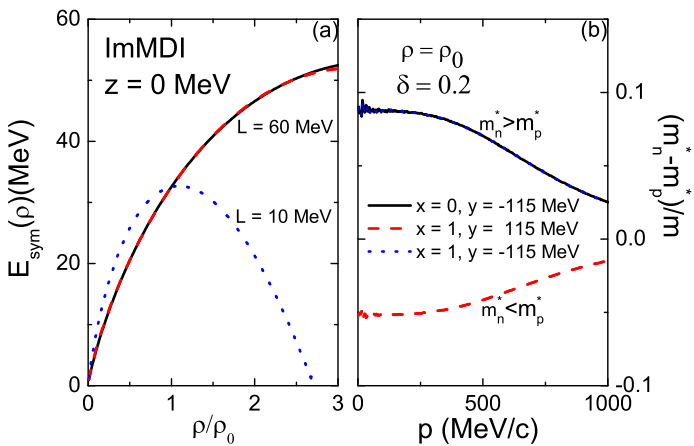}}
\caption{(Color online) Density dependence of the symmetry energy
(a) and momentum dependence of the relative neutron-proton effective
mass splitting in nuclear matter of the density $\rho=\rho_0$ and
isospin asymmetry $\delta=0.2$ (b) from the ImMDI interaction with
different values of $x$ and $y$. } \label{F1}
\end{figure}

The ImMDI interaction provides us with more flexibility to
investigate the detailed isovector properties of nuclear
interaction. In the present work, we set $z=0$ and vary
the values of $x$ and $y$ to study effects of the symmetry
energy and the neutron-proton effective mass splitting. The
nucleon effective mass is defined as
\begin{equation}
\frac{m_{\tau }^{\ast }}{m}=\left( 1+\frac{m}{p}\frac{dU_{\tau
}}{dp}\right) ^{-1}, \label{Meff}
\end{equation}%
and it generally depends on the density and isospin asymmetry of the
medium as well as the isospin and momentum of the nucleon. The
density dependence of the symmetry energy $E_{sym}$ and the relative
neutron-proton effective mass splitting are displayed in
Fig.~\ref{F1} with different values of $x$ and $y$. As discussed and
shown in Ref.~\cite{ImMDI}, $x$ affects only $E_{sym}$ while $y$
affects both $E_{sym}$ and the $m^*_{n-p}$. In the following, we select several special sets of parameters
to examine the relative effects of $E_{sym}$ and $m^*_{n-p}$.
With ($x=0$, $y=-115$ MeV) and ($x=1$, $y=115$ MeV),
the symmetry energy is almost the same while the relative
neutron-proton effective mass splittings are opposite in sign. On the other hand,
with ($x=0$, $y=-115$ MeV) and ($x=1$, $y=-115$ MeV) the relative
neutron-proton effective mass splitting is the same while the latter
gives a softer symmetry energy. We can thus study the effect of the
isospin splitting of nucleon effective mass by comparing the results
from the former two parameters sets while investigate that of the
symmetry energy by comparing the results from the latter two sets. We note that the current uncertainty range of the slope parameter $L$ of $E_{sym}$ is about $50 \pm 20$ MeV~\cite{Che12}, which is not quite different from the chosen range (10, 60) MeV in the present study.

The ImMDI interaction with parameter sets described above was implemented in the IBUU11
model~\cite{Li08}. In our calculation, 200 test
particles per nucleon are used and about 20,000 events are generated for each
beam energy and impact parameter. The initial density distribution
is generated from the Skyrme-Hartree-Fock calculation using the MSL0
force~\cite{Che10}, and the initial nucleon momenta are generated from local
Thomas-Fermi approximation.


\begin{figure}[h]
\centerline{\includegraphics[scale=0.8]{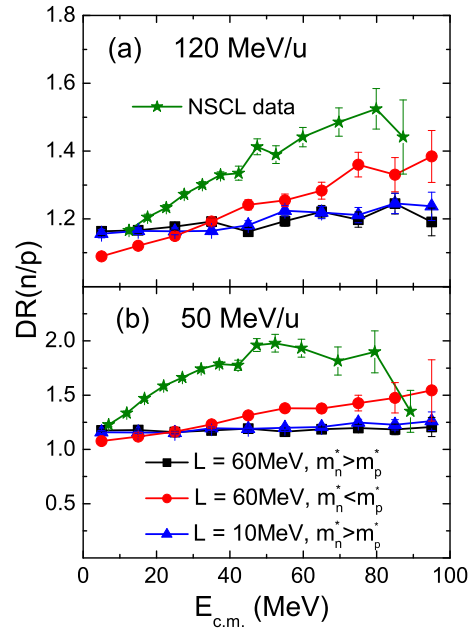}} \caption{(Color
online) The coalescence invariant double neutron/proton ratios
DR(n/p) in $^{124}$Sn+$^{124}$Sn collisions to $^{112}$Sn+$^{112}$Sn
collisions as a function of nucleon center-of-mass energy at beam
energies of 120 (a) and 50 MeV/u (b) with the impact parameter
$\text{b}=2$ fm and the angular gate
$70^{\circ}<\theta_{cm}<110^{\circ}$. The NSCL data are from
Ref.~\cite{Cou14}.} \label{F2}
\end{figure}

The neutron/proton ratio in collisions induced by
neutron-rich nuclei was firstly used as a probe of the symmetry
energy~\cite{Li97}. Later, the double neutron/proton ratio of
nucleon emission for two collision systems with isotopes of
different total isospin asymmetries was introduced to reduce
systematically the influence of the Coulomb force and the poor
efficiency of detecting neutrons~\cite{Li06,Kum11}. To explore the effects
of both the symmetry energy and the isospin splitting of nucleon
effective mass on double neutron/proton ratio within the ImMDI and
IBUU framework, we generate events for $^{112}$Sn+$^{112}$Sn and
$^{124}$Sn+$^{124}$Sn collisions at beam energies of 120 and
50 MeV/u. Similar to the treatment in Ref.~\cite{Li97}, we stop the
evolution at $t=150$ fm/c when the interaction becomes negligible,
and identify nucleons and clusters based on the final nucleon
phase-space distribution, i.e., two nucleons are within one cluster
if their spatial distance is closer than $\Delta r=3$ fm and their
momentum distance is smaller than $\Delta p=300$ MeV/c. We notice that our final results are
not sensitive to the variation of these coalescence parameters within about $30\%$ of the above values.

The coalescence invariant yield is constructed by considering both free
nucleons and those bound in light clusters including deuterons, tritons, $^{3}$He,
and $^{4}$He, and the angular gate is chosen to be
$70^{\circ}<\theta _{cm}<110^{\circ}$, as in the experimental
analysis. The impact parameter is set to be $\text{b}=2$ fm to mimic
the centrality in the experiments. The double neutron/proton ratio
DR(n/p) in $^{124}$Sn+$^{124}$Sn and $^{112}$Sn+$^{112}$Sn collisions
is defined as
\begin{equation}
\text{DR(n/p)}=\frac{\left[Y(n)/Y(p)\right]_{^{124}\rm {Sn}+^{124}\rm {Sn}}}{\left[Y(n)/Y(p)\right]_{^{112}\rm {Sn}+^{112}\rm {Sn}}}.
\end{equation}
Since the yield neutron/proton ratio $Y(n)/Y(p)$ in
$^{124}$Sn+$^{124}$Sn collisions is larger than that in
$^{112}$Sn+$^{112}$Sn collisions, the DR(n/p) is always larger than 1.
Figure~\ref{F2} shows the DR(n/p) as a function of nucleon
center-of-mass energy at beam energies of 120 and 50 MeV/u.
At 50 MeV/u, it is seen that the DR(n/p) for
high-energy nucleons is slightly larger for a softer symmetry energy
($L=10$ MeV) consistent with the finding in Ref.~\cite{Li06}. It is noteworthy that a
stiffer symmetry energy can lead to a larger DR(n/p) at beam energies
as high as 400 MeV/u, when the symmetry energy at suprasaturation
densities becomes important. At the beam energy of 120 MeV/u shown in panel (a) of
Fig.~\ref{F2}, the DR(n/p) is rather insensitive to the stiffness of
the symmetry energy. On the other hand,  the neutron-proton effective mass splitting has a more appreciable effect on the DR(n/p).
It is seen that a negative $m^*_{n-p}$  results in a larger DR(n/p) at higher nucleon energies, consistent
with our expectation. However, even for the two extreme cases considered here, the resulting values of the
DR(n/p) are still far below the NSCL data. We notice that in
the above calculation the isospin-dependent in-medium
nucleon-nucleon scattering cross sections scaled by the nucleon
effective mass~\cite{Li05} are used, and we have checked that the
results do not change by much even if we use free nucleon-nucleon
scattering cross sections. Thus, within the present IBUU11 transport model,
the variation of neither the symmetry energy nor the isospin effective mass splitting within their current uncertainty ranges
is able to explain the experimental data. This situation clearly calls for possibly new mechanisms and explanations to resolve the neutron/proton ratio puzzle.
It is thus interesting to note that several microscopic many-body theories and phenomenological models have predicted that the isospin dependence of short-range nucleon-nucleon
correlations dominated by the tensor force can significantly reduce the kinetic part of the symmetry energy \cite{XuLi,Vid11,Lov11,Car12,Rio14}. The potential part of the symmetry energy and thus the
symmetry potential has to be enhanced accordingly to meet existing constraints on the symmetry energy at saturation density. This effect has been shown to enhance significantly the double neutron/proton ratio
of free nucleons in transport model calculations without considering the neutron-proton effective mass splitting~\cite{Hen14,LiBA14}.  A comprehensive study considering the isospin dependence of short-range
nucleon-nucleon correlations and the neutron-proton effective mass splitting in the IBUU11 model is planed.

Trusting the data, the apparent success of the ImQMD model with the SLy4 interaction and the failure of the IBUU11 model with the ImMDI interaction in describing the data requires further investigations. Generally speaking,  different inputs and algorithms used in different transport models may lead to different predictions. We speculate, that the different handing of cluster formation might contribute appreciably to the difference between the ImQMD and IBUU11 calculations. However, we honestly do not know at this time what are really the main causes because each model has several major inputs besides some technical differences. One possible way out of this unfortunate situation is to conduct multi-observables versus multi-inputs covariance analyses, which have been successfully utilized in several other areas of nuclear physics recently (see Ref.~\cite{JPG} for a topic review). In the covariant analyses, the correlation matrix among all observables and input parameters as well as their uncertainties can be calculated consistently and simultaneously. Such analyses with the IBUU11 model are underway. It will also be useful to perform such analyses with the ImQMD model. For example, the ImQMD calculations using the SLy4 and SkM* interactions predict an approximately $50\%$ difference in the free n/p ratio at 50 MeV/u. Where does this difference come from? It was attributed to the difference in the isospin effective mass splitting in Ref.~\cite{Cou14}. In fact, at $\delta=0.2$, $m^*_n/m_n$ and $m^*_p/m_p$ with SLy4 are respectively 0.68 and 0.71, while that with SkM* are respectively 0.82 and 0.76, with relative effective mass splitting $-3\%$ for SLy4 and $+6\%$ for SkM*, smaller than that in the present calculation. However, these are not the only differences between the two interactions. In particular, the isoscalar effective mass is different by about $14\%$, and the curvature of the symmetry energy $K_{sym}$ differs by $30\%$. Without examining sensitivities of observables to a particular input by varying it while fixing all others, it is hard to conclude what input is actually responsible for the observed change in any observable. Thus, multi-dimenstonal covariance analyses look promising although they are computationally extremely costly using transport models for nuclear reactions.

In summary, within the IBUU11 transport model using a newly improved isospin- and momentum-dependent interaction, we revisited but failed to resolve the neutron/proton ratio puzzle in heavy-ion collisions at intermediate energies. Nevertheless, some interesting physics and useful lessons are learned.  We found that the neutron-proton effective mass splitting $m^*_n-m^*_p$ indeed has a relatively stronger effect than the symmetry energy $E_{sym}(\rho)$ and the assumption of $m^*_n\leq m^*_p$ leads to a higher neutron/proton ratio of free nucleons and those in light clusters.
Using the $E_{sym}(\rho)$ and $m^*_n-m^*_p$ both within their current uncertainty ranges, with the IBUU11 model and the ImMDI interaction we are unable to reproduce the recent NSCL/MSU double neutron/proton ratio data in central $^{124}$Sn+$^{124}$Sn and $^{112}$Sn+$^{112}$Sn collisions at 50 and 120 MeV/u. This situation clearly calls for new mechanisms to explain the puzzlingly high
neutron/proton ratio observed in the experiments. Among the possible new physics origins, effects of the isospin-dependent short-range nucleon-nucleon correlation deserve special attention.

We thank W.G. Lynch, M.B. Tsang, Y.X. Zhang, and W.J. Guo for
helpful communications, and C. Zhong for maintaining the high-quality
performance of the computer facility. This work was supported in part by the Major
State Basic Research Development Program (973 program) in China
under Contract Nos. 2015CB856904, 2014CB845401, and 2013CB834405,
the National Natural Science Foundation of China under Grant Nos.
11475243, 11421505, 11275125, 11135011, and 11320101004, the
"100-talent plan" of Shanghai Institute of Applied Physics under
Grant No. Y290061011 from the Chinese Academy of Sciences, the
"Shanghai Pujiang Program" under Grant No. 13PJ1410600, the "Shu
Guang" project supported by Shanghai Municipal Education Commission
and Shanghai Education Development Foundation, the Program for
Professor of Special Appointment (Eastern Scholar) at Shanghai
Institutions of Higher Learning, the Science and Technology
Commission of Shanghai Municipality (11DZ2260700), the US National
Science Foundation grants PHY-1068022, and the CUSTIPEN (China-U.S. Theory
Institute for Physics with Exotic Nuclei) under DOE grant number
DE-FG02-13ER42025.

\end{document}